\documentclass[prl,twocolumn,superscriptaddress,showpacs,amsmath,amssymb]{revtex4}
\usepackage{graphicx}
\usepackage{indentfirst}
\usepackage{psfrag}
\usepackage{epsfig}
\usepackage{amsmath}
\usepackage{amssymb}
\usepackage{bm}

\begin{document}

\title{Creation of Van der Waals, Casimir, and many more stochastic forces, with light radiation pressure via optics of randomly fluctuating sources} 
\author{Manuel Nieto-Vesperinas and Juan Miguel  Au\~n\'{o}n (2) }
\affiliation{Instituto de Ciencia de Materiales de Madrid, Consejo Superior de
Investigaciones Cient\'{i}ficas\\
 Campus de Cantoblanco, Madrid 28049, Spain.\\ www.icmm.csic.es/mnv; 
mnieto@icmm.csic.es \\
(2) School of Physics and Astronomy and School of Biology and Medicine
University of St Andrews, Scotland, U.K.}




\maketitle

The mechanical action on matter of the electromagnetic field emitted by a fluctuating source is governed by its statistics. In particular, thermal sources and vacuum fluctuations exert on bodies those well-known Casimir (C) and Van der Waals (VdW) forces \cite{Milonni, Hankel, Antezza, Capasso} .  

However, we have recently demonstrated that partially coherent random electromagnetic fields emitted by tailored optical sources, induce a photonic force on particles which, in particular, may be equivalent to those of Van der Waals and Casimir.  The reason is that vacuum fluctuation forces may be optically created on taking into account that the working optical frequencies $\omega$  are such that

\begin{equation}\hbar\omega/kT\gg1, \end{equation}
at $T=300K$, and hence Planck energy becomes the same as that of the vacuum fluctuations:

\begin{equation} \hbar\omega[\frac{1}{2}+1/(\exp(\hbar\omega/kT)-1)]\approx \frac{1}{2}\hbar\omega. \end{equation}
Hence, if we considered the emitting optical source spectrum as just given by a Planck distribution, the forces in the visible and NIR ranges due to the particle induced fluctuating dipoles, will be the optical analogous to those from the vacuum fluctuations in the thermal spectrum. 

In the same way, optical forces with random light at optical frequencies  can be tailored to be similar to  Lifshitz forces from thermal radiation at finite temperature. Thus an optical system constitutes an excellent means to create, test and monitoring photonic analogous of such thermal forces as well as of those out of thermodynamic equilibrium .

This leads to a two-fold important consequence: 

(i) Because of the high control of measurements in optical set-ups, designing partially coherent wavefields leads to a deeper understanding of the conditions and limitations under which theories and detections of these forces are established. In this way, one can accurately study the light-induced interaction, and in particular that of C and VdW,  between particles and surfaces \cite{Nieto1}, or in  the optical binding between particles \cite{Nieto2}. 

(ii) The possibilities of creating arbitrary random forces through optics of controlled partially coherent fields are  far-reaching beyond thermal effects from blackbodies. One does not have to restrict to Planck-like broad spectra, but can design the bandwidth and coherence length of the emitted field at will. For instance, photonic forces of nature similar to those of C and VdW can be studied with partially coherent quasimonochromatic sources acting on magnetodielectric particles, which have a rich spectral behavior and so much interest have aroused because of their potential in harvesting light energy, or as  metamaterial elements and directional nanoantennas.

\begin{figure}[h]
\begin{centering}
\includegraphics[width=\linewidth]{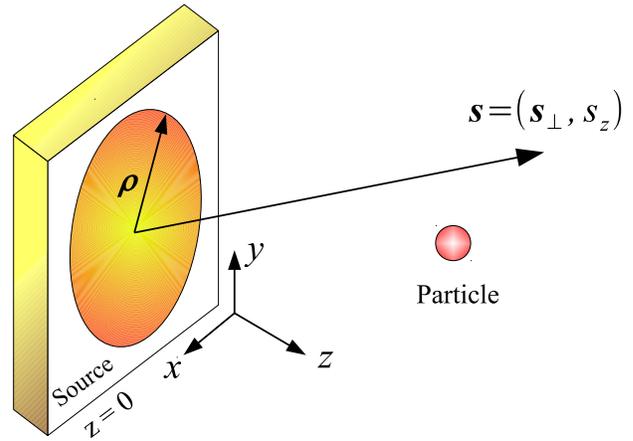}
\par\end{centering}
\caption{Scheme  to create Casimir, Van der Waals, and other partially coherent photonic forces: A  beam, either quasimonochromatic or polychromatic, passes through a phase screen, or fluctuating slab, of designed power spectrum and coherence length, that acts as the random planar source. The emitted field in directions ${\bf s}$ induces those  forces on particles in its vicinity.}
\end{figure}

\end{document}